# Impact of Facebook Usage on Undergraduate Students Performance in Irbid National University: Case Study

**Fawzi H. Altaany, Firas A. Jassim**
Department of Management Information Systems, Irbid 2600, Jordan

**ABSTRACT**
The aim of this study is to investigate the style of Facebook usage between undergraduate students and the impact on their academics performance. Also, this paper was evaluated in the view of student the using of Facebook. A questioner was design for collecting data from a sample of 480 undergraduate students in Irbid National University. The survey revealed that 77% of the students have an account on Facebook. One of the main findings is that there was a significant relationship between gender and Facebook usage. Moreover, the survey revealed that whenever the less time spent on Facebook, the higher the performance will be in grade point average. This was conducted by the negative correlation between time spent on Facebook and the performance of undergraduate student. Statistically speaking, the study has seven hypotheses; two of them were rejecting against five acceptable hypotheses.

## I.  INTRODUCTION

Many tools online are creative for communication among people, but the creatively and amazing one is Facebook which connects people to gathers throughout world [1]. The emergence of technology world and network developed the use of Facebook, the aim or objective of Facebook in any universities developed relational, friendly or social network among students and their collage [2]. Many studies stated or argued that Facebook participation enhances connection between the students and social communication among them [3].

Students use Facebook for social network, however friendship, relational with other student to do their project, discussion material, and others thing like games [4]. Nowadays, Facebook is the most famous site on internet, and appeared as the result of technology. Facebook was innovated and created in February 2004 as a site for Harvard students only. But with years ago Facebook was deployed as a social network like other social websites. Facebook has become popular among collage students and affect their performance, either positively or negatively. Facebook offers much option for users as a chat option, away for communication and to write messages. Facebook was a dialogue, discussion and exchange of views among students. Users of Facebook make then add their friend to chat or receive their messages or sent them; people may make groups with others who are interested in the same subject.

Nowadays, more than 500 million user use Facebook with wide social network, Ellison et al 2007 stated that 94% of the university students use Facebook. According to [5], monthly active users reaches nearly 850 million, photos are uploading every day. Moreover, 20% of all pages that were viewed on the web were viewed by Facebook, 425 million mobile users and 100 billion connections. Furthermore, using games revenue is currently 12% of Facebook with total income 2.7 billion. In accordance with gender, 57% of users are females.

This research is the first research of its kind in Irbid National University of undergraduate students. The aim of this study is to know the level of effectiveness of Facebook on the performance of undergraduate students in Irbid National University. Also, the aim of this paper is to find the time spent by students in using Facebook and what is the correlation between performances of students as average accumulated or grade and the time on Facebook spent.

The aimed of this study is to investigate the following:
1. To identify style of Facebook usage between students.
2. To know the effect of Gender on Facebook usage.
3. To evaluate the frequent usage of Facebook among students
4. To compute academic performance within the usage of Facebook.
5. To know students the interest of performance enhancing with Facebook.

## II.  LITERATURE REVIEW

According to [6], studied the Facebook usage and their academic performance in public and private universities, the result of their study shows that the greater time spent by students on Facebook, the lower the academic performance. [4] studied pattern of Facebook use and academic performance. Their result was that the more time spent on Facebook, the lower academic performance of students. But this result was mainly affected by male students rather than females. In accordance with [3], a study was established to test the influence of Facebook usage on the academic performance and the quality of life of college students. The result indicates that smoking have affected the performance of students and their life, but the males grades were lower than female students and males experience in Facebook usage was better than females. Moreover, an interesting study for Facebook usage was researched in university of Karlstad for 595 users had found that undergraduate male students spent more time on Facebook rather than female students [1].

It must be mentioned that, there is negatively relationship between Facebook and academic performance, however the more activity and time spent (at least once a day) on Facebook usage the lower the academic performance [7]. In reference to [8], study relationship among frequency of Facebook usage and engagement, stated that there is a non-positive correlation of Facebook engagement with time spend. In another study, [9] tested relationship among many indices of Facebook usage and academic performance, the result of the study shows that more time spent on Facebook has negative correlation with grade point average of students. From research of university population [6][4][10][11], the following hypotheses were made:

1. Less time spending on Facebook, higher grade point average student performance.
2. Most of the students in the university use Facebook.
3. Students in the university use Facebook for communication in their academic material achievement will be positive with grade point average performance.
4. Greater time spent on Facebook leads to more friends.
5. There is a significant relationship between gender and Facebook usage with friend.
6. Males in the university spent more time on Facebook more than females.
7. Fourth and third year students spent more time than first and second year students in the university.

### III. METHOD AND RESULTS

Researchers have focused on impact of usage Facebook on undergraduate students in Irbid National University in Jordan and their performance. ANOVA, correlations and frequencies used to test the hypotheses in this study.

3.1. Results:
Random sample of (480), (301) male and (179) female undergraduate students were selected from Irbid National University. The levels of academics range are: first year (114) second year (83), third year (99) and fourth (184).

The survey revealed that 77% of the students have an account on Facebook, 23% did not have an account on Facebook, from the 23% of the students who does not have an account on Facebook 27.3% does not know Facebook, 72.7% does not care about Facebook. The average of students was accept 102 (21.3%), good 155 (32.3), very good 146 (30.4%) and excellent 77 (16%). The Facebook accounts that were created using real names were 307 (83%) and using nickname were 63 (17%). The number of hours students spent daily on Facebook was less than one hours 151 (40.8%), from one to less than two hours 89 (24.1%), two to less than three hours 55 (14.9%) and more than three hours 75 (20.3%). The daily usage Facebook was 257 (69.5%), once weekly was 48 (13%), only once monthly 17 (4.6) and more than month 48 (13%). The percentages of student who felt that Facebook affects the average were 310 (65.7%) of students strongly disagree and disagree while 162 (34.3%) strongly agree and agree. The percentages of student who felt that Facebook usage is harmful were 285 (60.4%) of students strongly disagree and disagree while 187 (39.6%) strongly agree and agree. Moreover, another question was tested which is that does Facebook support your academic study? The answer was 338 (70.4%) of students strongly disagree and disagree while 134 (29.6%) strongly agree and agree. Another test question was: does Facebook usage is on of the reasons of your absence? The answer was 348 (70.4%) strongly disagree and disagree while 134 (29.6%) strongly agree and agree. The student were use Facebook with friends 353 (74.5), while 127 (26.5) of students use Facebook with non-friends. In general the students were asked about Facebook usage for long time? The answer was that more than 4 hours daily is 361 (75.2%) strongly disagree and disagree while 119 (24.8%) strongly agree and agree.

### IV. EXAMINE THE HYPOTHESES

Hypothesis one: Less time spending on Facebook, higher grade point average student performance will be. The dependant variable was the average of student (acceptable, good, very good and excellent), the predictors are does Facebook affect your absence, have an account on Facebook, number of hours spent on Facebook, how long (years) on Facebook, academic level and Facebook usage. Table (1) illustrates the significant of the effect of predictor variables on average student performance. We accept this hypothesis because it's significant (0.000) which is less than (0.05).

The second hypothesis: most of the students in Irbid National University use Facebook. Have an account on Facebook. To test this hypothesis we asked question which is that do you have a Facebook account, the answer was either yes or no. we accept this hypothesis because the results revealed that 77% of the students have an account on Facebook, 23% did not have an account on Facebook, from percent 23% between student does not have an account on Facebook 27.3% does not know Facebook, 72.7% does not care about Facebook. The result of hypothesis was shown in table (2).

Third hypothesis: students in Irbid National University use Facebook for communication of academic material achievement will be positive with grade point average performance. As a result of analysis in tables (3, 4, 5) there is a significant relationship between these variables but the correlation was a negative, this mean we reject this hypothesis.

The fourth hypothesis: greater time on Facebook spent by the students in Irbid National University, leads to more friends gained. Table (6), revealed that the correlation between Facebook usage with friend and number of hours spent on Facebook, the table indicates a positive relationship and significant (0.000), the correlation was (0.227). Because of this result we accept this hypothesis.

Hypothesis five: there is a significant in Irbid National University between gender and Facebook usage with friend. In table (7, 8) of model summary and ANOVA illustrates a weak relationship but it was significant with (0.000), in this hypothesis we accept it.

Sixth hypothesis: males in Irbid National University spent more time on Facebook than females. From table (9), the total male on Facebook usage was (247) 67% and the for females was (123) 33%, number of hours spent by males on Facebook for less than one hour was (61%), between one to two hours was (88%), between two to three hours was (24%), more than three hours (85%) and the female number of hours spent on face book for less than one hour was (39%), between one to two hours was (22%), between two to three hours was (76%) and more than three hours (15%), this means that males spent more time than females, we accept the hypothesis.

The seventh hypothesis: Final year students in Irbid National University spent more time than intermediate and first year students. In table (10), the correlation between academic levels and number of hours spent on Facebook was (- 0.144) and with significant (0.006), this mean that high levels students spent more time rather than intermediate and first year students. We reject this hypothesis.

Table (1)

**ANOVA$^b$**

| Model | | Sum of Squares | df | Mean Square | F | Sig. |
|---|---|---|---|---|---|---|
| 1 | Regression | 37.473 | 6 | 6.245 | 7.574 | .000$^a$ |
| | Residual | 299.338 | 363 | .825 | | |
| | Total | 336.811 | 369 | | | |

a. Predictors: (Constant), Does facebook affect your absence?, Have an account on facebook?, Number of hours spent on facebook?, How long in Facebook, Academic level, facebook usage

b. Dependent Variable: Average

Table (2)

**Have an account on facebook?**

| | | Frequency | Percent | Valid Percent | Cumulative Percent |
|---|---|---|---|---|---|
| Valid | Yes | 371 | 77.3 | 77.3 | 77.3 |
| | No | 109 | 22.7 | 22.7 | 100.0 |
| | Total | 480 | 100.0 | 100.0 | |

Table (3)

**Model Summary**

| Model | R | R Square | Adjusted R Square | Std. Error of the Estimate |
|---|---|---|---|---|
| 1 | .162[a] | .026 | .024 | .98305 |

a. Predictors: (Constant), Does facebook help you in your HW?

Table (4)

**ANOVA[b]**

| Model | | Sum of Squares | df | Mean Square | F | Sig. |
|---|---|---|---|---|---|---|
| 1 | Regression | 12.395 | 1 | 12.395 | 12.827 | .000[a] |
|   | Residual | 461.930 | 478 | .966 | | |
|   | Total | 474.325 | 479 | | | |

a. Predictors: (Constant), Does facebook help you in your HW?

b. Dependent Variable: Average

Table (5)

**Correlations**

| | | Average | Does facebook help you in your HW? |
|---|---|---|---|
| Average | Pearson Correlation | 1 | -.162** |
|  | Sig. (2-tailed) |  | .000 |
|  | N | 480 | 480 |
| Does facebook help you in your HW? | Pearson Correlation | -.162** | 1 |
|  | Sig. (2-tailed) | .000 |  |
|  | N | 480 | 480 |

**. Correlation is significant at the 0.01 level (2-tailed).

Table (6)

**Correlations**

| | | How long in Facebook | Number of hours spent on facebook? | facebook usage | Usage of facebook with freinds |
|---|---|---|---|---|---|
| How long in Facebook | Pearson Correlation | 1 | -.044 | -.172** | .087 |
|  | Sig. (2-tailed) |  | .395 | .001 | .062 |
|  | N | 464 | 370 | 370 | 464 |
| Number of hours spent on facebook? | Pearson Correlation | -.044 | 1 | -.434** | .227** |
|  | Sig. (2-tailed) | .395 |  | .000 | .000 |
|  | N | 370 | 370 | 370 | 370 |
| facebook usage | Pearson Correlation | -.172** | -.434** | 1 | -.142** |
|  | Sig. (2-tailed) | .001 | .000 |  | .006 |
|  | N | 370 | 370 | 370 | 370 |
| Usage of facebook with freinds | Pearson Correlation | .087 | .227** | -.142** | 1 |
|  | Sig. (2-tailed) | .062 | .000 | .006 |  |
|  | N | 464 | 370 | 370 | 480 |

**. Correlation is significant at the 0.01 level (2-tailed).

Table (7)

**Model Summary**

| Model | R | R Square | Adjusted R Square | Std. Error of the Estimate |
|---|---|---|---|---|
| 1 | .177[a] | .031 | .029 | .47691 |

a. Predictors: (Constant), Usage of facebook with freinds

Table (8)

**ANOVA[b]**

| Model | | Sum of Squares | df | Mean Square | F | Sig. |
|---|---|---|---|---|---|---|
| 1 | Regression | 3.532 | 1 | 3.532 | 15.530 | .000[a] |
| | Residual | 108.716 | 478 | .227 | | |
| | Total | 112.248 | 479 | | | |

a. Predictors: (Constant), Usage of facebook with freinds

b. Dependent Variable: Gender

Table (9)

**Gender * Number of hours spent on facebook? Crosstabulation**

Count

| | | Number of hours spent on facebook? | | | | Total |
|---|---|---|---|---|---|---|
| | | < hour | 1-2 hours | 2-3 hours | more than 3 hours | |
| Gender | Male | 92 | 78 | 13 | 64 | 247 |
| | Female | 59 | 11 | 42 | 11 | 123 |
| Total | | 151 | 89 | 55 | 75 | 370 |

Table (10)

**Correlations**

| | | Academic level | Average | How long in Facebook | Number of hours spent on facebook? | Have an account on facebook? | facebook usage | Does facebook affect your absence? |
|---|---|---|---|---|---|---|---|---|
| Academic level | Pearson Correlation | 1 | -.143** | .098* | -.144** | -.179** | .174** | -.117* |
| | Sig. (2-tailed) | | .002 | .034 | .006 | .000 | .001 | .010 |
| | N | 480 | 480 | 464 | 370 | 480 | 370 | 480 |
| Average | Pearson Correlation | -.143** | 1 | .235** | -.001 | -.025 | -.023 | -.188** |
| | Sig. (2-tailed) | .002 | | .000 | .987 | .587 | .652 | .000 |
| | N | 480 | 480 | 464 | 370 | 480 | 370 | 480 |
| How long in Facebook | Pearson Correlation | .098* | .235** | 1 | -.044 | -.478** | -.172** | -.184** |
| | Sig. (2-tailed) | .034 | .000 | | .395 | .000 | .001 | .000 |
| | N | 464 | 464 | 464 | 370 | 464 | 370 | 464 |
| Number of hours spent on facebook? | Pearson Correlation | -.144** | -.001 | -.044 | 1 | -.051 | -.434** | .095 |
| | Sig. (2-tailed) | .006 | .987 | .395 | | .324 | .000 | .067 |
| | N | 370 | 370 | 370 | 370 | 370 | 370 | 370 |
| Have an account on facebook? | Pearson Correlation | -.179** | -.025 | -.478** | -.051 | 1 | .118* | .121** |
| | Sig. (2-tailed) | .000 | .587 | .000 | .324 | | .023 | .008 |
| | N | 480 | 480 | 464 | 370 | 480 | 370 | 480 |
| facebook usage | Pearson Correlation | .174** | -.023 | -.172** | -.434** | .118* | 1 | -.157** |
| | Sig. (2-tailed) | .001 | .652 | .001 | .000 | .023 | | .002 |
| | N | 370 | 370 | 370 | 370 | 370 | 370 | 370 |
| Does facebook affect your absence? | Pearson Correlation | -.117* | -.188** | -.184** | .095 | .121** | -.157** | 1 |
| | Sig. (2-tailed) | .010 | .000 | .000 | .067 | .008 | .002 | |
| | N | 480 | 480 | 464 | 370 | 480 | 370 | 480 |

**. Correlation is significant at the 0.01 level (2-tailed).

*. Correlation is significant at the 0.05 level (2-tailed).

## V. CONCLUSION

The scope of this paper composed via Facebook and the common usage for undergraduate students. The result of this study was to indicate a number of significant implications to grasp why and how undergraduate students used Facebook. The study was operating to reduce the negative effects and increase the positive effects of Facebook usage. The study has seven hypotheses; two of them were rejected, five hypotheses were accepted. Most of the students in Irbid National University were using Facebook. The students were use Facebook with friend. The time spent on Facebook were significant with the average of performance, less time spent on Facebook higher of average; this may due to Facebook usage with friends for a along time. The study revealed that Facebook usage for communication of academic material achievement had a negative relationship with grade point average performance. Because in this study most of the student use Facebook with friends and do not use it with homework, there is a significant relationship in gender about Facebook usage with friends and time spent.